\documentclass[10pt,twocolumn,oneside,conference]{IEEEtran}
\usepackage{cite}
\usepackage{graphicx}
\usepackage{amsmath}
\usepackage{times}
\usepackage{latexsym}
\usepackage{graphicx}
\usepackage{bm}
\usepackage{amssymb}
\usepackage[center]{caption2}
\usepackage{stfloats}
\usepackage{cases}
\usepackage{array}
\usepackage{setspace}
\usepackage{fancyhdr}
\usepackage{algorithm}
\usepackage{algorithmic}

\usepackage{amsmath,amsthm}
\usepackage{graphicx}
\usepackage{subfigure}

\usepackage{xcolor}

\usepackage{balance} 
\usepackage{flushend} 

\allowdisplaybreaks[4]

\newcaptionstyle{mystyle1}{%
  \centering TABLE  \captiontext \par}
\captionstyle{mystyle1}
\newcaptionstyle{mystyle2}{%
  \captionlabel $.$ \: \captiontext \par}
\captionstyle{mystyle2}
\newcaptionstyle{mystyle3}{%
  \captionlabel $.$ \: \captiontext \par}
\captionstyle{mystyle3}

\begin{document}

\title{Pattern Division Multiple Access \\
with Large-scale Antenna Array}



\author{
\IEEEauthorblockN{Peng Li$^\dag$, Yanxiang Jiang$^{\dag*}$, Shaoli Kang$^\ddag$, Fuchun Zheng$^\dag$$^\S$, and Xiaohu You$^\dag$}
\IEEEauthorblockA{$^\dag$National Mobile Communications Research Laboratory,
Southeast University, Nanjing 210096, China.\\
$^\ddag$China Academy of Telecommunication
Technology, Beijing 100080, China.\\
$^\S$Department of Electronics, University of York, UK.\\
$^*$E-mail: yxjiang@seu.edu.cn
}}

\maketitle

\begin{abstract}

    In this paper, pattern division multiple access with large-scale antenna array (LSA-PDMA) is proposed as a novel non-orthogonal multiple access (NOMA) scheme.
    In the proposed scheme, pattern is designed in both beam domain and power domain in a joint manner.
    At the transmitter, pattern mapping utilizes power allocation to improve the system sum rate and beam allocation to enhance the access connectivity and realize the integration of LSA into multiple access spontaneously.
    At the receiver, hybrid detection of spatial filter (SF) and successive interference cancellation (SIC) is employed to separate the superposed multiple-domain signals.
    Furthermore, we formulate the sum rate maximization problem to obtain the optimal pattern mapping policy,
    and the optimization problem is proved to be convex through proper mathematical manipulations.
    Simulation results show that the proposed LSA-PDMA scheme achieves significant performance gain on system sum rate compared to both the orthogonal multiple access scheme and the power-domain NOMA scheme.

\end{abstract}

\section{Introduction}

    With the new challenges of explosive mobile data growth, tremendous increment in the number of connected devices, and the continuous emergence of new service requirements, future 5G communications systems are emerging.
    In order to efficiently support unprecedented requirements for system sum rate and access connectivity, researchers from both industry and academia are focusing on non-orthogonal multiple access (NOMA) and large-scale antenna array (LSA) technologies \cite{5G, MassiveMIMO,overview_LSAS_2016}.

    In mobile communications systems, the design of multiple access schemes is of great importance to improve the system sum rate in a cost-effective manner.
    Actually, NOMA schemes are optimal in the sense of achieving the capacity region of the broadcast channel \cite{NOMA}.
    In NOMA schemes, multi-user signals are superposed in the same frequency and time resource blocks via code domain and/or power domain multiplexing at the transmitter, and separated at the receiver by multi-user detection based on successive interference cancellation (SIC), message passing algorithm (MPA) or maximum likelihood algorithm (MLA).
    Recently, several representative NOMA schemes have been proposed, such as power-domain NOMA (pNOMA), sparse code multiple access (SCMA) and pattern division multiple access (PDMA).
    pNOMA was introduced in \cite{pNOMA} using superposition coding at the transmitter and SIC at the receiver, which lays the foundation for NOMA schemes.
    SCMA was proposed in \cite{SCMA1}, where bit streams are directly mapped to sparse codewords, and thus it is amenable  to the  use of MPA with acceptable complexity \cite{SCMA2}.
    Different from the above mentioned NOMA schemes, PDMA adopts pattern segmentation to separate user signals at the transmitter and SIC at the receiver \cite{PDMA1, PDMA3}.
    However, a complete scheme of transmitter and receiver based on PDMA has not been reported yet.
    Besides, there has been little research work on the design of NOMA scheme in multiple domains other than power domain.

    On the other hand, as one of the key technologies of 5G mobile communications systems, LSA has been put forward to significantly improve the spectrum efficiency with extra degrees of freedom which facilitate transmit diversity and spatial multiplexing gains \cite{MassiveMIMO}.
    Facing massive amounts of connected devices, LSA can provide sufficient spatial resource.
    More recently, the application of LSA to NOMA has been receiving growing attention for further capacity improvement \cite{NOMAmassiveMIMO1, NOMAmassiveMIMO2}.

    Motivated by the aforementioned results, in this paper, we propose a PDMA scheme with LSA (LSA-PDMA).
    In the proposed scheme, pattern is designed in both beam domain and power domain in a joint manner.
    Pattern mapping at the transmitter utilizes power allocation and beam allocation to superpose user signals, while hybrid detection of spatial filter (SF) and SIC is  employed at the receiver to separate the superposed multiple-domain signals.
    Besides, we investigate the optimal pattern mapping policy.
    By exploiting the convexity of the sum rate maximization problem, the globally optimal pattern mapping policy can be readily obtained.
    Our major contributions are summarized as follows:
    \begin{itemize}
    \item {We propose to use beams in space domain as the multiplexing resource shared by the users, and thus the integration of LSA into multiple access can be realized spontaneously, which makes it significantly different compared to the other NOMA schemes.}
    \item {Furthermore, pattern mapping at the transmitter not only enhances the access connectivity greatly, but also reduces the computational load of the receiver.} 
    \item {Most importantly, the proposed scheme can be considered as a superset of pNOMA scheme and orthogonal multiple access (OMA) scheme: it can easily be transformed to the latter schemes by adjusting the pattern mapping policy.}
    \end{itemize}

    The rest of the paper is organized as follows.
    The system model is briefly presented in Section II.
    The proposed LSA-PDMA scheme is described in Section III.
    Simulation results are shown in Section IV.
    Final conclusions are drawn in Section V.

\section{System Model}

\begin{figure}[!t]
\centering
\includegraphics[width=0.38\textwidth]{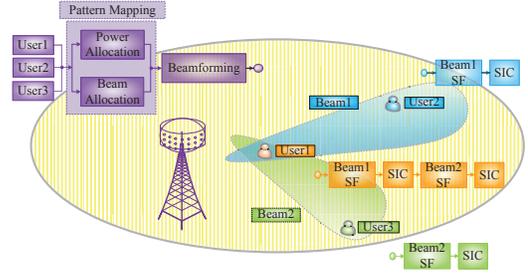}
\caption{Illustration of the proposed LSA-PDMA scheme.}
\label{scheme}
\end{figure}

    In this paper, we consider a downlink transmission scenario with one base station (BS) communicating with multiple users.
    {The BS is equipped with LSA, and each user has \(N_R\) antennas.
    Assume that there are multiple antenna clusters (AC) located in the BS and each AC equipped with \(N_T\) antennas forms \(N\) beams, where \(N_T \geq N\).
    All the users are divided into multiple user groups (UG) and each UG contains \(K\) users, where \(N_T \le KN_R\).
    Assume that an AC covers a UG with \(N \le K \le 2^N-1\) \cite{PDMA1}.
    Without loss of generality, we simplify the scenario into the case where an AC communicates with a UG as illustrated in Fig. \ref{scheme}.}


    Let \(\boldsymbol{G}_k \in {\mathbb{C}^{N_R \times N_T}}\) denote the channel matrix between the AC and the \(k\)-th user in the UG.
    Assume that zero-forcing beamforming (ZFBF) is utilized at the transmitter.
    Let ${\boldsymbol{f}_n \in {\mathbb{C}^{N_T \times 1}}}$ denote the ZFBF vector of the \(n\)-th beam, and it is generated based on the CSI of the selected user among the users covered by the \(n\)-th beam denoted with the index $n_{\rm{sel}}$.
    And the set of pairs of beam and the corresponding selected user is expressed as \(\Omega  = \left\{ (1\text{, }{1_{\rm{sel}}})\text{, }(2\text{, }{2_{\rm{sel}}})\text{, } \cdots \text{, }(n\text{, }{n_{\rm{sel}}}) \cdots \text{, }(N\text{, }{N_{\rm{sel}}}) \right\}\).
    Then, the composite channel matrix of the selected users can be expressed as \({\boldsymbol{G}_C} = {\left[ {\boldsymbol{G}_{1_{\rm{sel}}}^H\text{, }\boldsymbol{G}_{2_{\rm{sel}}}^H\text{, } \cdots \text{, }\boldsymbol{G}_{N_{\rm{sel}}}^H} \right]^H}\), and the composite ZFBF matrix can be expressed as \(\boldsymbol{F}_C = \boldsymbol{G}_C^H{\left( {\boldsymbol{G}_C \boldsymbol{G}_C^H} \right)^{ - 1}}\).
    We partition the composite ZFBF matrix as \(\boldsymbol{F}_C = \left[ {\boldsymbol{F}_1\text{, } \boldsymbol{F}_2\text{, } \cdots \text{, } \boldsymbol{F}_n\text{, } \cdots \text{, } \boldsymbol{F}_N} \right]\), and then the ZFBF vector of the \(n\)-th beam can be expressed as follows
\begin{equation}\begin{split}\label{ZFBF}
{\boldsymbol{f}_n} = {\boldsymbol{F}_n}{{\bf{1}}^{N_R \times 1}}\text{,}
\end{split}\end{equation}
    where \({\bf{1}}^{N_R \times 1}\) denotes an \(N_R\)-dimension column vector with all one elements.
    Therefore, let $\boldsymbol{F} \in {\mathbb{C}^{N_T \times N}}$ denote the ZFBF matrix and it can be expressed as \(\boldsymbol{F}=\left[ {\boldsymbol{f}_1\text{, }\boldsymbol{f}_2\text{, } \cdots \text{, } \boldsymbol{f}_n\text{, } \cdots \text{, } \boldsymbol{f}_N} \right]\).

    At the transmitter, let \(\boldsymbol{t} \in {\mathbb{C}^{N \times 1}}\) denote the superposed signal vector after the process of pattern mapping.
    Let \(\boldsymbol{x} \in {\mathbb{C}^{N_T \times 1}}\) denote the transmit signal vector from the AC.
    Then, it can be expressed as follows
\begin{equation}\begin{split}\label{transmitted_signal}
\boldsymbol{x} = \boldsymbol{F} \boldsymbol{t}\text{.}
\end{split}\end{equation}

    At the receiver, let \(\boldsymbol{y}_k \in {\mathbb{C}^{N_R \times 1}}\) denote the received signal vector for the \(k\)-th user.
    Then, it can be expressed as follows
\begin{equation}\begin{split}\label{received_signal}
{\boldsymbol{y}_k} = {\boldsymbol{G}_k}\boldsymbol{x} + {\boldsymbol{w}_k}\text{,}
\end{split}\end{equation}
    where \(\boldsymbol{w}_k \sim \mathcal{CN}(\bold{0}\text{, } \sigma_k^2{\bold{I}_{N_R})}\) is the additive white Gaussian noise vector whose elements have zero mean and variance $\sigma_k^2$.

    In following, pattern mapping is designed to generate the superposed signal vector \(\boldsymbol{t}\) and hybrid detection is proposed to process the received signal vector \(\boldsymbol{y}_k\).
    And they are jointly designed based on both beam domain and power domain.

\section{The Proposed LSA-PDMA Scheme}

    In the following, we will elaborate on  pattern mapping at the transmitter and hybrid detection at the receiver.
    Furthermore, the optimization of pattern mapping will be discussed.

\subsection{Pattern Mapping at the Transmitter}

    The multiple-domain multiplexing and superposition coding are the key factors of pattern design in PDMA.
    In general, power domain is chosen as the basis in the pattern design \cite{PDMA1, PDMA3}.
    However, in the case of single power domain, the stringent requirement in power allocation policy restricts its application to various scenarios other than the macro cell.
    The combination of multiple domains can make the most of wireless resource, and generalize PDMA to various application scenarios.

    In this paper, pattern of the LSA-PDMA scheme is designed based on the combination of power domain and beam domain.
    By considering LSA, multiple beams in downlink can serve as spatial resource blocks.
    Specifically, beams are shared by multiple users with different power, and the specific allocation policy depends on pattern mapping.

    In Fig. \ref{scheme}, pattern mapping at the transmitter utilizes power allocation and beam allocation to superpose multiple-domain signals.
    Let \(\boldsymbol{s} \in {\mathbb{C}^{K \times 1}}\) denote the transmit symbol vector for the UG with \(\boldsymbol{s} \sim \mathcal{CN}(\bold{0}\text{, }\bold{I}_K)\), \({\left[ \boldsymbol{s} \right]_{k}} = {s_{k}}\) the transmit symbol for the \(k\)-th user.
    Let \(\boldsymbol{P}^{1/2} \in {\mathbb{R}^{N \times K}}\) denote the power allocation matrix, \({\left[ \boldsymbol{P}^{1/2} \right]_{nk}} = {\sqrt {p_{nk}}}\) the transmit power allocated to the \(k\)-th user in the \(n\)-th beam.
    Let \(\boldsymbol{B} \in {{[0,1]}^{N \times K}}\) denote the beam allocation matrix, \({\left[ \boldsymbol{B} \right]_{nk}} = {b_{nk}}\) the beam allocation factor for the \(k\)-th user in the \(n\)-th beam with \({b_{nk}} \in \left\{ {0\text{, }1} \right\}\).
    Specifically, the factor $b_{nk}=1$ means that the $k$-th user is covered by the $n$-th beam.
    Correspondingly, the superposed signal vector after pattern mapping can be expressed as follows
\begin{equation}\begin{split}\label{superposedSignal}
\boldsymbol{t} = \left( {\boldsymbol{B} \circ {\boldsymbol{P}}^{1/2}} \right) \boldsymbol{s}\text{,}
\end{split}\end{equation}
    where \({\left[ \boldsymbol{t} \right]_{n}} = t_n = \sum\limits_{k = 1}^K {{b_{nk}}\sqrt {{p_{nk}}} {s_k}} \), and $\circ$ denotes the Hadamard product.

\begin{figure}[!tbp]
\centering
\includegraphics[width=0.38\textwidth]{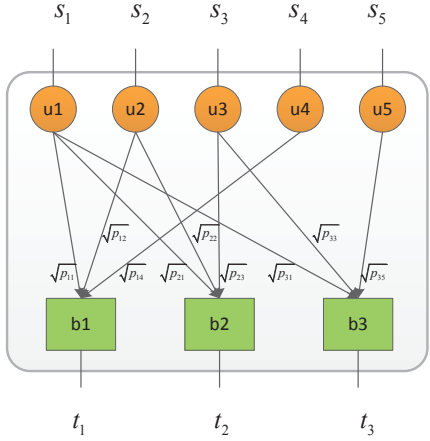}
\caption{The factor graph corresponding to pattern mapping with \(N=3\) and \(K=5\).}
\label{mappingmatrix}
\end{figure}

    To further analyze the structure of pattern mapping, we assume that \(\boldsymbol{B}\) with \(N=3\) and \(K=5\) is designed as follows
\begin{equation}\begin{split}\label{B35}
{\boldsymbol{B}_{3 \times 5}} = \left[ {\begin{array}{*{20}{c}}
1&1&0&1&0\\
1&1&1&0&0\\
1&0&1&0&1
\end{array}} \right]\text{,}
\end{split}\end{equation}
    whose factor graph is shown in Fig. \ref{mappingmatrix}.
    And more details on the optimal structure of pattern mapping can be found in the subsection C in the Section II.
    Specifically, the factor graph shows the mapping relationship from \(\boldsymbol{s}\) to \(\boldsymbol{t}\).
    Each row of \(\boldsymbol{B}\) denotes a beam, and each column of \(\boldsymbol{B}\) denotes an user.
    Correspondingly, the diversity in pattern mapping refers to the sum number of the corresponding column in \(\boldsymbol{B}\) for the considered user, and the overlap in pattern mapping refers to the sum number of the corresponding row in \(\boldsymbol{B}\) for the considered beam.
    There are some insights in the structure of pattern mapping.

    Firstly, we remark here that the proposed LSA-PDMA scheme can be considered as a superset of pNOMA scheme and OMA scheme: it can easily be transformed to the latter schemes by adjusting the diversities and overlaps in pattern mapping.
    By setting all the diversities to be one, the proposed LSA-PDMA scheme can be transformed to pNOMA scheme.
    While it can be transformed to OMA scheme by setting all the diversities and overlaps to be one.

    Furthermore, let $\lambda$ denote the overload ratio supported by pattern mapping and it can be expressed as $\lambda=K/N$.
    By exploiting pattern mapping, access connectivity in the LSA-PDMA scheme can thus be enhanced up to a maximum of \(\left( {{2^N} - 1} \right)/N\) folds.

    Finally, note here that the maximum diversity in pattern mapping should be $N$ \cite{PDMA1}, i.e., there are certain user transmits power in all beams.
    Therefore, in the case of narrow beamwidth, the maximum number of beams in one AC covering the corresponding UG is not large.
    Specifically, a small matrix $\boldsymbol{B}$ in pattern mapping is the common case for the proposed LSA-PDMA scheme.
    And beams are chosen as spatial resource in the proposed scheme other than antennas or space-time block code \cite{PDMA2} in space domain.
    {Therefore, when combined with LSA, the dimension size of $\boldsymbol{B}$ is reduced from ${N_T} \times K$ to $N \times K$ with \(N_T \geq N\), which can reduce the computational load for the receiver.}

\subsection{Hybrid Detection of SF and SIC at the Receiver}

     For the received signals, hybrid detection is proposed as illustrated in Fig. \ref{scheme}.
     Specifically, SF is performed to suppress the inter-beam interference caused by beam-domain multiplexing, and SIC is performed to remove the intra-beam interference caused by power-domain multiplexing.

     For the received signal vector, SF is first performed to suppress the inter-beam interference.
     Let \({\boldsymbol{V}_{k}} \in {\mathbb{C}^{N_R \times N}}\) be the SF matrix for the \(k\)-th user, \({\boldsymbol{v}_{nk}}\) the \(n\)-th column of \({\boldsymbol{V}_{k}}\).
     Assume that \({\boldsymbol{V}_{k}}\) is calculated based on the minimum mean square error (MMSE) criteria \cite{pNOMA2} as follows
\begin{equation}\begin{split}\label{SF}
{\boldsymbol{V}_{k}} & = \mathop {\min }\limits_{{\boldsymbol{\tilde V}_k}} {\mathbb E}\left\{ {\left\| {\boldsymbol{t} - \boldsymbol{\tilde V}_k^H{\boldsymbol{y}_k}} \right\|_2^2} \right\} \\
& = {\left( {{\boldsymbol{G}_k}\boldsymbol{FA}{\boldsymbol{F}^H}\boldsymbol{G}_k^H + \sigma _k^2{{\bf{I}}_{N_R}}} \right)^{ - 1}}{\boldsymbol{G}_k}\boldsymbol{FA}\text{,}
\end{split}\end{equation}
    where \(\boldsymbol{A} \buildrel \Delta \over = \mathbb{E}\left\{ {\boldsymbol{t}{\boldsymbol{t}^H}} \right\}\), whose element can be expressed as \({\left[ \boldsymbol{A} \right]_{ij}} = \sum\limits_{k = 1}^K {{b_{ik}}{b_{jk}}\sqrt {{p_{ik}}{p_{jk}}} } \).
    By considering the matrix inversion involved in (\ref{SF}), the computational complexity of $\left( {{\boldsymbol{G}_k}\boldsymbol{FA}{\boldsymbol{F}^H}\boldsymbol{G}_k^H + \sigma _k^2{{\bf{I}}_{N_R}}} \right)^{-1}$ is approximately $O(N_R^3)$.
    In a real scenario, $N_R$ is always very small.
    Therefore, the complexity of SF is reasonable.

    Let \(z_{nk}\) denote the received signal after the process of SF, and it can be expressed as follows
\begin{equation}\begin{split}\label{scalar_received_signal}
{z_{nk}} = \boldsymbol{v}_{nk}^H{\boldsymbol{y}_k}{\rm{ = }}\boldsymbol{v}_{nk}^H{\boldsymbol{G}_k}{\boldsymbol{f}_n}{t_n} + \boldsymbol{v}_{nk}^H{\boldsymbol{G}_k}\sum\limits_{i=1\hfill\atop i\ne n\hfill}^N {{\boldsymbol{f}_i}{t_i}}  + \boldsymbol{v}_{nk}^H{\boldsymbol{w}_k}
\end{split}\end{equation}
    where the first term of the right hand denotes the combination of the desired information and intra-beam interference, while the other terms denote the inter-beam interference and noise, respectively.
    Let the aggregated power of the inter-beam interference and noise in \(z_{nk}\) be normalized to be one.
    And let \(h_{nk}\) denote the equivalent normalized channel gain between the \(k\)-th user and the AC.
    Then, it can be expressed as follows
\begin{equation}\begin{split}\label{h_nk}
{h_{nk}} = \sqrt {\frac{{{{\left| {\boldsymbol{v}_{nk}^H{\boldsymbol{G}_k}{\boldsymbol{f}_n}} \right|}^2}}}{{\sum\limits_{i = 1\hfill\atop
i \ne n\hfill}^N {{{\left| {\boldsymbol{v}_{nk}^H{\boldsymbol{G}_k}{\boldsymbol{f}_i}} \right|}^2}}  + \sigma _k^2{{\left\| {{\boldsymbol{v}_{nk}}} \right\|}^2}}}}\text{.}
\end{split}\end{equation}
    Correspondingly, the expression in (\ref{scalar_received_signal}) can be reshaped as follows
\begin{equation}\begin{split}\label{normalized}
{z_{nk}} = {h_{nk}}\sum\limits_{i = 1}^K {{b_{ni}}\sqrt {{p_{ni}}} {s_i}}  + {q_{nk}}\text{,}
\end{split}\end{equation}
    where \(q_{nk}\) represents the sum of the inter-beam interference and noise after normalization with \(\mathbb{E}\left[ {{{\left| {{q_{nk}}} \right|}^2}} \right] = 1\).
    Therefore, the multiple-input multiple-out (MIMO) channel between the \(k\)-th user and the AC can degrade into a single-input single-out (SISO) channel after the normalization \cite{pNOMA2}, which meets the implementation condition of SIC.

    For the scalar received signal, SIC is employed at the receiver to remove the intra-beam interference.
    The key idea of SIC is to decode symbols iteratively by subtracting the detected symbols of strong users first to facilitate the detection of weak users.
    Without loss of generality, we assume that the \(K\) users are placed in an ascending order of normalized channel gain \(h_{nk}\) with respect to their index numbers.
    For instance, \({h_{ni}} \le {h_{nj}}\) holds if \(1 \le i \le j \le K\).
    Consequently, the \(j\)-th user can correctly decode the signal symbol in spite of the interference of the \(i\)-th user.
    As a result, the signal to interference plus noise ratio (SINR) of the \(k\)-th user at the \(n\)-th beam can be expressed as follows
\begin{equation}\begin{split}\label{SINR}
{\gamma _{nk}} = {\rm{ }}\left\{ {\begin{array}{*{20}{l}}
{\frac{{h_{nk}^2{b_{nk}}{p_{nk}}}}{{1 + h_{nk}^2\sum\limits_{i = k + 1}^K {{b_{ni}}{p_{ni}}} }}\text{, } \ k = 1\text{, } \cdots \text{, }K - 1}\text{, }\\
{h_{nk}^2{b_{nk}}{p_{nk}}\text{, } \ k = K}\text{.}
\end{array}} \right.
\end{split}\end{equation}
    Then, the sum rate of the AC can be expressed as follows
\begin{equation}\begin{split}\label{sumrate}
{R_{{\rm{sum}}}} = \sum\limits_{n = 1}^N {\sum\limits_{k = 1}^K {{{\log }_2}\left( {1 + {\gamma _{nk}}} \right)} }\text{.}
\end{split}\end{equation}


\subsection{Optimization of Pattern Mapping Policy}

    To further improve the system performances, the optimization of pattern mapping at the transmitter is investigated here.

    At the transmitter, the beam allocation matrix ${\boldsymbol{B}}$ can be embodied within the power allocation matrix ${\boldsymbol{P}}^{1/2}$.
    ${\sqrt{p_{nk}}=0}$ implies that ${b_{nk}=0}$ and the $n$-th beam is not shared by the $k$-th user.
    ${\sqrt{p_{nk}}>0}$ implies that ${b_{nk}=1}$ and the $n$-th beam is shared by the $k$-th user with transmit power ${\sqrt{p_{nk}}}$.
    Therefore, we adopt ${\boldsymbol{\tilde P}}$ to denote the pattern mapping matrix which combines ${\boldsymbol{B}}$ and ${\boldsymbol{P}}^{1/2}$, and ${\left[{\boldsymbol{\tilde P}}\right]}_{nk} = {\tilde p_{nk}} = {b_{nk}}{p_{nk}}$.
    Then, the expression in (\ref{SINR}) can be reshaped as follows
\begin{equation}\begin{split}\label{SINR2}
{\gamma _{nk}} = {\rm{ }}\left\{ {\begin{array}{*{20}{l}}
{\frac{{h_{nk}^2{\tilde p_{nk}}}}{{1 + h_{nk}^2\sum\limits_{i = k + 1}^K {{\tilde p_{ni}}} }}\text{, } \ k = 1\text{, } \cdots \text{, }K - 1}\text{, }\\
{h_{nk}^2{\tilde p_{nk}}\text{, } \ k = K}\text{.}
\end{array}} \right.
\end{split}\end{equation}

    Correspondingly, the optimization of pattern mapping at the transmitter can be simplified as follows:
\begin{equation}\begin{split}\label{OP}
\mathop {\max }\limits_{\boldsymbol{\tilde P}} \mathop {}\nolimits^{} \quad & \mathop {R_{\text{sum}}} = \sum\limits_{n = 1}^N {\sum\limits_{k = 1}^K {{{\log }_2}\left( {1 + {\gamma _{nk }}} \right)} } \\
\text{s.t. } \quad & \text{C1 : } {\tilde p_{nk}} \ge {\delta _{nk}}\text{, } \forall n\text{, } \forall k\text{, } \\
\quad & \text{C2 : } \sum\limits_{n = 1}^N {\sum\limits_{k  = 1}^K {{\tilde p_{nk}}} }  \le {P_{{\text{sum}}}}\text{, } \\
\quad & \text{C3 : } {\log _2}\left( {1+{\gamma_{nk}}} \right) \ge {R_{{\text{min}}}}\text{, } \forall n\text{, } \forall k\text{, }
\end{split}\end{equation}
    where ${\delta _{nk}} = \left\{ {\begin{array}{*{20}{l}} {\varepsilon\text{, } \text{if} \ (n\text{, }k) \in \Omega } \\ {0\text{, } \text{if} \ (n\text{, }k) \notin \Omega } \end{array}} \right.$, $\varepsilon$ denotes the slack variable for the selected user of ZFBF, ${P_{{\text{sum}}}}$ denotes the maximum sum transmit power for the AC, and ${R_{{\text{min}}}}$ denotes the minimum rate requirement for the \(k\)-th user at the \(n\)-th beam.
    {We remark here that the pattern mapping problem is degraded into a power allocation problem when ${R_{{\text{min}}}>0}$, where each user transmits power in all the beams.}
    Then, we have the following theorem.

    \textit{Theorem 1:} {The optimization problem in (\ref{OP}) is a convex problem.}

    \textit{ \ Proof:} {Please refer to the Appendix.}
    $\hfill{} \blacksquare$

    To solve the convex optimization problem in (\ref{OP}), we adopt the barrier method \cite{Book} to get the globally optimal pattern mapping policy.

\section{Simulation Results}


    In this section, we compare the performances of our proposed LSA-PDMA scheme with the OMA scheme and the pNOMA scheme.
    The numbers of antennas in the BS and each user are set as \(N_T=16\) and \(N_R=4\), respectively.
    The BS is located in the cell center with radius 800 meters.
    It is assumed that all the users in the considered UG are distributed uniformly inside the cell. 
    Considering the propagation channel, it is assumed that the complex propagation coefficient between each antenna of the BS and each antenna of each user is modeled as a complex small-scale fading factor timed by a large-scale fading factor, which models geometric attenuation and shadow fading.
    For the small-scale fading factor, it is always assumed to be i.i.d. random variable with distribution \(\mathcal{CN}(0\text{, }1)\).
    For the large-scale fading factor, the path loss factor, the path loss exponent and the variance of log-normal shadow fading are set to be 1, 3.7 and 10dB, respectively.

    Assume that the considered AC in the BS contains \(N=3\) beams, unless otherwise stated.
    As for the OMA scheme, the number of users is set to be 3, i.e., each user monopolizes a beam.
    As for the pNOMA scheme, the number of users is set to be 6, i.e., every 2 users share a beam \cite{NOMA}.
    As for the LSA-PDMA scheme, the number of users is from 3 to 7, which means the proposed LSA-PDMA scheme can perform under the varying overload ratio (\(\lambda = 100\% \sim 233\%\)).

    The performance of the LSA-PDMA scheme is firstly evaluated by adopting a simple pattern mapping policy.
    In this policy, fixed-ratio power allocation \cite{pNOMA} and simple beam allocation \cite{PDMA1} are adopted.
    For the fixed-ratio power allocation, users are sorted in an ascending order of the normalized channel gain.
    Let \(p_0\) denote the basic transmit power, which is allocated to the first scheduled user.
    And then the transmit power allocated to the \(k\)-th scheduled user is set to be \({\mu ^{k-1}}{p_0}\), where \(\mu\) denotes the power gain factor.
    {For the simple beam allocation}, the basic principle is that larger diversity is allocated to the user with smaller normalized channel gain, and vice versa.

\begin{figure}[!tbp]
\centering
\includegraphics[width=0.38\textwidth]{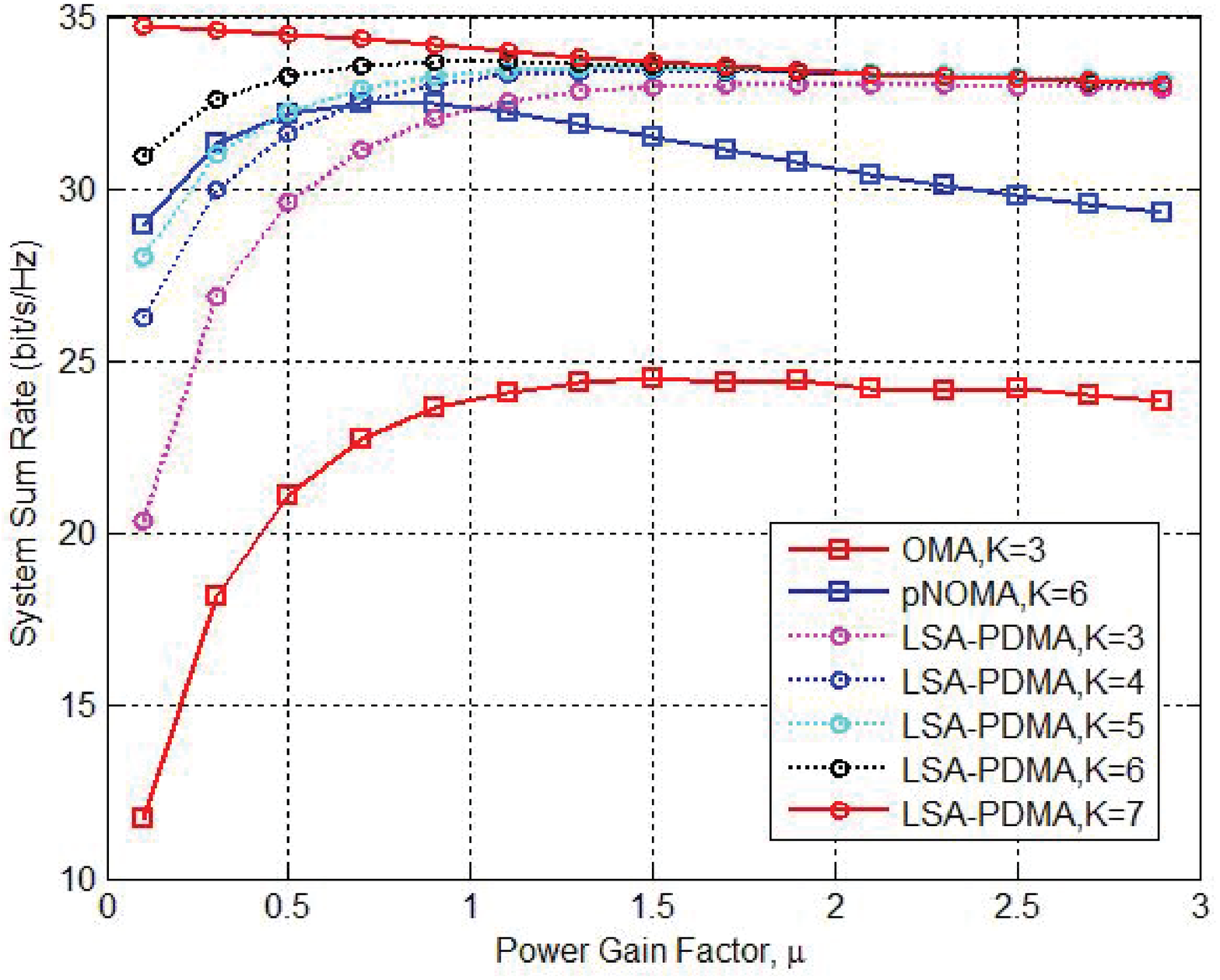}
\caption{The system sum rate vs. the power gain factor, where the simple pattern mapping policy is adopted for the LSA-PDMA scheme.}
\label{powergain}
\end{figure}

    Fig. \ref{powergain} depicts the system sum rate versus the power gain factor where the simple pattern mapping policy is adopted for the proposed scheme.
    The sum transmit power is set to be \(10\text{dB}\) here.
    When $\mu>1$, more power is allocated to the stronger user. 
    It can be observed that the performances of the LSA-PDMA scheme stop increasing and tend to be constant when \(\mu\) is larger than a certain threshold.
    This results from the design principles of beamforming.
    In the proposed LSA-PDMA scheme, beamforming is generated based on the CSI of the weakest user within a beam for achieving the user fairness.
    Therefore, the detection performance of the stronger user at the receiver is poorer, and  the stronger user contributes less to the system sum rate even with more transmit power.
    When $\mu<1$, more power is allocated to the weaker user. 
    It can be observed that the impact of smaller \(\mu\) is reduced when the overload ratio increases.
    The reason is that the weaker user allocated more power under the case of smaller \(\mu\) offsets the performance loss.
    When the overload ratio becomes larger, the weaker user contributes more to the system performance.
    Furthermore, it can also be observed that the performance gain of the LSA-PDMA scheme even with the simple pattern mapping policy is significant compared with the OMA scheme and the pNOMA scheme.

\begin{figure}[!tbp]
\centering
\includegraphics[width=0.38\textwidth]{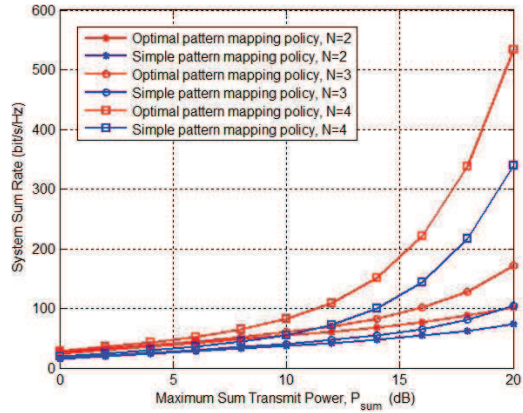}
\caption{The system sum rate vs. the maximum sum transmit power $P_\text{sum}$.}
\label{cvx}
\end{figure}

    In Fig. \ref{cvx}, we compare the system sum rate of the proposed LSA-PDMA scheme adopting different pattern mapping policies for \(N=2 \text{, }3\text{, }4\), and $K=2^N-1$.
    It is obvious that the optimal pattern mapping policy achieves the remarkable performance gain over the simple pattern mapping policy.
    And the more resource can be allocated, the greater performance gain can be achieved by the optimal pattern mapping policy.

\begin{figure}[!tbp]
\centering
\includegraphics[width=0.38\textwidth]{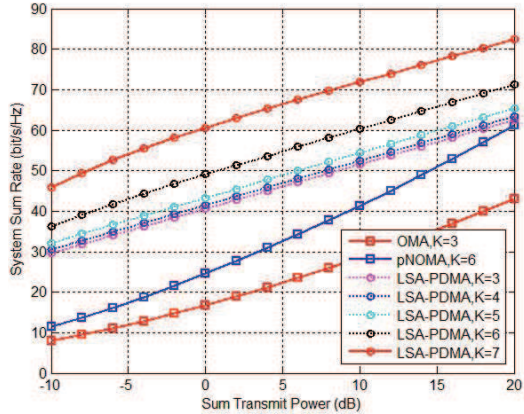}
\caption{The system sum rate vs. the sum transmit power, where the optimal pattern mapping policy is adopted for the LSA-PDMA scheme.}
\label{sim10s}
\end{figure}

    Fig. \ref{sim10s} shows how the sum transmit power allocated to the UG affects the system sum rate, where the optimal pattern mapping policy is adopted for the LSA-PDMA scheme.
    It can be observed that the performances of all the schemes improve as the sum transmit power increases.
    It can also be observed that our proposed scheme significantly outperforms the OMA scheme and pNOMA scheme.
    Besides, we can see that the performance of the LSA-PDMA scheme gets better when the overload ratio becomes larger.
    And the performance gain is not significant enough when the overload ratio is less than \(200\%\).
    The reason is that the gain from multiple-domain multiplexing cannot offset the loss caused by the intra-beam interference in SIC with the small overload ratio.

\newcounter{TempEqCnt}
\setcounter{TempEqCnt}{\value{equation}}
\setcounter{equation}{2}
\begin{figure*}[!b]
\hrulefill
\vspace*{4pt}
\setcounter{equation}{13}
\begin{equation}\begin{split}\label{2pd}
{\nabla ^2}f{\left( {{\boldsymbol{\tilde p}_n}} \right)_{ij}} = \left\{ {\begin{array}{*{20}{l}}
{\frac{1}{{{{\left( {\frac{1}{{h_{n1}^2}} + \sum\limits_{l = 1}^K {{\tilde p_{nl}}} } \right)}^2}\ln 2}}\text{, }   {i = 1\text{, }\forall j}  \ \text{or}  \ {j = 1\text{, }\forall i}  ,}\\
{\frac{1}{{{{\left( {\frac{1}{{h_{n1}^2}} + \sum\limits_{l = 1}^K {{\tilde p_{nl}}} } \right)}^2}\ln 2}} + \frac{1}{{\ln 2}}\sum\limits_{l = 1}^{\min \left( {i\text{, }j} \right) - 1} {\left( {\frac{1}{{{{\left( {\frac{1}{{h_{n(l + 1)}^2}} + {w_{nl}}} \right)}^2}}} - \frac{1}{{{{\left( {\frac{1}{{h_{nl}^2}} + {w_{nl}}} \right)}^2}}}} \right)}\text{, } i \ne 1\text{, } j \ne 1.}
\end{array}} \right.
\end{split}\end{equation}
\setcounter{equation}{\value{TempEqCnt}}
\end{figure*}


\section{Conclusions}

    In this paper, we have designed a joint LSA-PDMA scheme based on  both joint beam domain and power domain.
    The proposed scheme realized the integration of LSA into multiple access spontaneously.
    Furthermore, the LSA-PDMA scheme can be considered as a superset of pNOMA and OMA schemes.
    Even with the simple pattern mapping policy, the proposed scheme can achieve significant performance gain.


\vspace{-5pt}

\vspace{-5pt}

\appendix
\section*{Proof of Theorem 1}
    \textit{ \ Proof:}
    The optimization problem in (\ref{OP}) is firstly reshaped as a standard form problem \cite{Book} as follows
\begin{equation*}\begin{split}\label{OP2}
\quad & \mathop {\min }\limits_{\boldsymbol{\tilde P}} \mathop {}\nolimits^{} \mathop f{\left( \boldsymbol{\tilde P} \right)} = - \sum\limits_{n = 1}^N {\sum\limits_{k = 1}^K {{{\log }_2}\left( {1 + \frac{{h_{nk}^2{\tilde p_{nk}}}}{{1 + h_{nk}^2w_{nk} }}} \right)} } \\
\text{s.t.} \quad & \text{C1:} g_1^{nk}{\left( \boldsymbol{\tilde P} \right)} = {\delta_{nk}} - {\tilde p _{nk}} \le 0\text{, } \forall n\text{, } \forall k\text{, } \\
\quad & \text{C2:} g_2{\left( \boldsymbol{\tilde P} \right)} = \sum\limits_{n = 1}^N {\sum\limits_{k  = 1}^K {{\tilde p_{nk}}} } - {P_{{\text{sum}}}} \le 0\text{, } \\
\quad & \text{C3:} g_3^{nk}{\left( \boldsymbol{\tilde P} \right)} = {R_{{\text{min}}}} - {\log _2}\left( {1+\frac{{h_{nk}^2{\tilde p_{nk}}}}{{1 + h_{nk}^2w_{nk} }}} \right) \le 0\text{, } \\
\quad & {\rm{ }} \  \   \  \quad \quad\quad\quad\quad\quad\quad  \quad\quad\quad\quad \quad \quad\quad\quad\quad\quad\quad \forall n\text{, } \forall k\text{, }
\end{split}\end{equation*}
    where
${w_{nk}} = \left\{ {\begin{array}{*{20}{l}}
{\sum\limits_{i = k + 1}^K {{\tilde p_{ni}}}\text{, }k = 1\text{, } \cdots \text{, }K - 1}\text{, }\\
{0\text{, }k  = K}\text{.}
\end{array}} \right.$

    It is obvious that the function $g_1^{nk}{\left( \boldsymbol{\tilde P} \right)}$ is affine in $\boldsymbol{\tilde P}$ for any $n$ and $k$, and the function $g_2{\left( \boldsymbol{\tilde P} \right)}$ is affine in $\boldsymbol{\tilde P}$ as well.
    Due to the independence of $h^2_{nk}$ and $w_{nk}$ on $\tilde p_{nk}$, the function $g_3^{nk}{\left( \boldsymbol{\tilde P} \right)}$ is convex in $\boldsymbol{\tilde P}$ for any $n$ and $k$.

    Now consider the objective function $f {\left( \boldsymbol{\tilde P} \right)}$.
    Due to the fact that $w_{nk}$ couples the multiple variables with respect to $k$ not $n$, the convexity of $f {\left( \boldsymbol{\tilde P} \right)}$ can be derived from the convexity of functions $f {\left( \boldsymbol{\tilde p}_n \right)} = - \sum\limits_{k = 1}^K {{{\log }_2}\left( {1 + \frac{{h_{nk}^2{\tilde p_{nk}}}}{{1 + h_{nk}^2{w_{nk}}}}} \right)}$, where ${\boldsymbol{\tilde p}_n} = \left[ {\tilde p}_{n1}\text{, } {\tilde p}_{n2}\text{, } \cdots \text{, } {\tilde p}_{nk}\text{, } \cdots \text{, } {\tilde p}_{nK} \right]$.

    Let ${\nabla^2}f{\left( {\boldsymbol{\tilde p}_n} \right)}$ denote the second derivative of $f{\left( {\boldsymbol{\tilde p}_n} \right)}$, whose element is given by (\ref{2pd}) at the bottom of this page.

    Let
\setcounter{equation}{15}
\begin{align*}
{\alpha_0} &\buildrel \Delta \over = \frac{1}{{{{\left( {\frac{1}{{h_{n1}^2}} + \sum\limits_{l = 1}^K {{\tilde p_{nl}}} } \right)}^2}\ln 2}} \text{,} \\
{\beta_m}& \buildrel \Delta \over = \frac{1}{{\ln 2}}\sum\limits_{l = 1}^{m} {\left( {\frac{1}{{{{\left( {\frac{1}{{h_{n(l + 1)}^2}} + {w_{nl}}} \right)}^2}}} - \frac{1}{{{{\left( {\frac{1}{{h_{nl}^2}} + {w_{nl}}} \right)}^2}}}} \right)} \text{.}
\end{align*}
    Recall the property of SIC: \({h_{ni}} \le {h_{nj}}\) holds if \(1 \le i \le j \le K\).
    Then, the term $\frac{1}{{{{\left( {\frac{1}{{h_{n(l + 1)}^2}} + {w_{nl}}} \right)}^2}}} - \frac{1}{{{{\left( {\frac{1}{{h_{nl}^2}} + {w_{nl}}} \right)}^2}}}$ is always nonnegative.
    We can get that $\alpha_0 > 0$ and $\beta_m \ge 0$ for \(1 \le m \le K-1\).

    Correspondingly, the Hessian matrix of $f \left( {{\boldsymbol{\tilde p}_n}} \right)$ can be expressed as follows
\begin{equation*}\begin{split}\label{hessian}
{\nabla^2}f{\left( {{\boldsymbol{\tilde p}_n}} \right)} =
\left[
\begin{array}{*{20}{c}}
{{\alpha_0}} & {{\alpha_0}} & {{\alpha_0}} & {\cdots} & {{\alpha_0}} \\
{{\alpha_0}} & {{\alpha_0}+{\beta_1}} & {{\alpha_0}+{\beta_1}} & {\cdots} & {{\alpha_0}+{\beta_1}} \\
{{\alpha_0}} & {{\alpha_0}+{\beta_1}} & {{\alpha_0}+{\beta_2}} & {} & {\vdots} \\
{\vdots} & {\vdots} & {} & {\ddots} & {} \\
{{\alpha_0}} & {{\alpha_0}+{\beta_1}} & {\cdots} & {} & {{\alpha_0}+{\beta_{K-1}}}
\end{array}
\right]\text{.}
\end{split}\end{equation*}
    Then, ${\nabla ^2}f\left( {{\boldsymbol{\tilde p}_n}} \right) \succeq {\bf{0}}$.
    Therefore, the function $f{\left( \boldsymbol{\tilde p}_n \right)}$ is convex in $\boldsymbol{\tilde p}_n$ for any $n$.
    According to the transitivity of the convexity \cite{Book}, the function $f{\left( \boldsymbol{\tilde P} \right)}$ is convex in $\boldsymbol{\tilde P}$ as well.

    The optimization problem in (\ref{OP}) is now proved to be convex.
    $\hfill{} \blacksquare$

\section*{Acknowledgments}

    This work was supported in part by the National 863 Project (2015AA01A709), the National Basic Research Program of China (973 Program 2012CB316004), the Natural Science Foundation of China (61221002, 61521061),
    and UK Engineering and Physical Sciences Research Council under Grant EP/K040685/2.

\end{document}